\documentclass{article}
\usepackage[T1]{fontenc}

\makeatletter

\newcommand{\LyX}{L\kern-.1667em\lower.25em\hbox{Y}\kern-.125emX\spacefactor1000}

\csname @addtoreset\endcsname{equation}{section}
\makeatother

\begin{document}

\begin{titlepage} 

\begin{flushright} {\small DFTT-58/98} \\ hep-th/9809145 \end{flushright} 

\vfill 

\begin{center} {

{\large The GS Type IIB Superstring Action on \( AdS_{3}\times S_{3}\times T^{4} \)} 

\vskip 0.3cm

{\large {\sl }} 

\vskip 10.mm 

{\bf I. Pesando }\footnote{e-mail: ipesando@to.infn.it, pesando@alf.nbi.dk\\

Work supported by the European Commission TMR programme ERBFMRX-CT96-004}

\vskip4mm

{\small Dipartimento di Fisica Teorica , Universit\'a di Torino, via P. Giuria 1, I-10125 Torino, Istituto Nazionale di Fisica Nucleare (INFN) - sezione di Torino, Italy} 

}

\end{center} 

\vfill

\begin{center}{ \bf ABSTRACT}\end{center} 

\begin{quote}

We describe the type IIB supergravity background on \( AdS_{3}\otimes S_{3}\otimes T^{4} \)using the potentials of \( AdS_{3|4}\otimes S_{3}\otimes T^{4} \) and we use the supersolvable algebra associated to \( AdS_{3|4} \) to compute the \( \kappa  \) gauge fixed type IIB string action. From the explicit form of the action we can clearly see how passing from pure NSNS backgrounds to pure RR backgrounds the WZW term disappears.


\vfill

\end{quote} 

\end{titlepage}

\section{Introduction.}

The AdS/CFT conjecture (\cite{Maldacena},\cite{GubserKlenanovPolyako},\cite{Witten})
has recently attracted a lot of interest and work, one of its consequences is
that the classical (i.e. no string loops) type IIB superstring propagating in
a \( AdS_{5}\otimes S_{5} \) background is the masterfield of the \( {\cal N}=4 \)
\( D=4 \) SYM based on the \( U(N) \) gauge group. It has therefore become
important to write the classical action for the above mentioned type IIB superstring
and then to first quantize it. The first step has been accomplished in a series
of works (\cite{MatsaevTseytlin},\cite{Kallosh},\cite{Kall-fixing}) which
culminated in two different proposal (\cite{Mio_ads},\cite{Kall_ads}) which
were shown to be equivalent (\cite{Mio_all}). Even if the resulting action
can be very easily written down and we can thrust it is the proper conformal
action (\cite{MatsaevTseytlin},\cite{kall-raj-vuoti}), it is unclear how to
proceed to quantize it explicitly, even in a simplified but equivalent form
(\cite{kall_ts-simpl}), the main reason being it describes a RR background
which reflects into the fact that the theory is not a usual WZW model because
it has not the left/right symmetry (\cite{MatsaevTseytlin}). In view of the
fact that explicit computations of \( {\cal N}=4 \) \( D=4 \) SYM amplitudes
in the leading order (\cite{4-punti}) and next to leading order (\cite{4-punti-corretta})
in \( \lambda  \) ( \( \lambda =Ng^{2}_{YM}>>1 \) ) based on this conjecture
have revealed the possible appearance of logarithm and even if the results in
this field are far from be definitive because of the difficulties of computing
all the contributions explicitly, it has become even more important to be able
to understand how to quantize this action.

In order to get some hint on how to proceed to quantize such a kind of action
we could start to look at some simpler example where one can at least partially
compare with a well defined NSR formulation: it is infact purpose of this paper
to write the explicit \( \kappa  \) gauge fixed action for type IIB string
propagating on \( AdS_{3}\otimes S_{3}\otimes T^{4} \) whose NSR formulation
has been worked out in (\cite{ads3-sei}). This non dilatonic background can
be pure NSNS, pure RR or a mixing of the two since it can be thought as generated
by a dionic string in the type IIB supergravity dimensional reduced to \( D=6 \)
: unfortunatly it turns out that the two backgrounds are so different to make
their comparison quite difficult.

The plan of the paper is the following.

In section 2 we compute the supergravitational background corresponding to a
spontaneous compactification on \( AdS_{3}\otimes S_{3}\otimes T^{4} \) and
we express the background fields, thought as fields living in a superspace,
through the fields of the superspace \( AdS_{3|4} \) and the vielbein and spin-connection
of \( S_{3}\otimes T^{4} \) .

In section 3 we derive the form of the background fields when fixing the \( \kappa  \)
symmetry using the supersolvable approach.

In section 4 we give the action for the type IIB superstring propagating in
this background, we discuss and comment the form of the action. 

In a series of appendiceswe give more details on the computations and we prove
that the proposed action actually fixes the \( \kappa  \) symmetry as claimed.

\section{The supergravitational background.}

We will now construct the supergravitational background explicitly: this will
be performed in three steps

\begin{enumerate}
\item we solve the equations of motion and we get the field strengths which are associated
with the spontaneous compactification on \( AdS_{3}\otimes S_{3}\otimes T^{4} \)
;
\item since the fermions are set to zero we look at their supersymmetric transformations
and we determine the form and the number of Killing vectors. Their form suggests
the ansatz for the gravitino in the next point;
\item we finally express all the fields through the generalized potentials of \( SU(1,1|2,0)\otimes SU(1,1|0,2) \)
and the vielbein and spin connection of \( S_{3}\otimes T^{4} \) : in this
way all the fields become defined on the superspace \( AdS_{3|4} \) .
\end{enumerate}

\subsection{The spontaneous compactification on \protect\( AdS_{3}\otimes S_{3}\otimes T^{4}\protect \)
.}

Since we want a solution of the equations of motion which has a spontaneous
compactification on \( AdS_{3}\otimes S_{3}\otimes T^{4} \), it is natural
to make the following ansatz which in the notation of \cite{IIB}\footnote{
With the minor modifications given by \( \psi \rightarrow \Psi  \) , \( \psi ^{*}\rightarrow \Psi _{c}\equiv \widehat{C}\overline{\Psi }^{T} \)and
\( \varpi ^{ab}\rightarrow \Omega ^{\widehat{a}\widehat{b}} \).
} reads

\begin{eqnarray}
R^{ab}_{..\: cd}=r_{1}\delta ^{ab}_{cd} & \; \; R^{ij}_{..\: kl}=r_{2}\delta ^{ij}_{kl} & \; \; R^{rs}_{..\: tu}=0\label{ansatz_eq_mot} \\
G_{abc}=g_{1}\epsilon _{abc} & \; \; \;  & G_{ijk}=g_{2}\epsilon _{ijk}\\
V^{\alpha }_{\pm }=const & \Rightarrow  & Q=0\; \; P_{a}=0\\
\;  & F_{\widehat{a}_{1}\ldots \widehat{a}_{5}}=0 & \; \\
\;  & \rho _{\widehat{a}\widehat{b}}=0 & \; \\
\lambda =0 & \Rightarrow  & D_{\widehat{a}}\lambda =0\label{ansatz_eq_mot_fine} 
\end{eqnarray}
where \( \widehat{a},\widehat{b},\ldots \in \{0,\ldots ,9\} \) , \( a,b,\ldots \in \{0,1,2\} \),
\( u,v,\ldots \in \{3,\ldots ,9\} \), \( i,j,\ldots \in \{3,4,5\} \), \( r,s,\ldots \in \{6,\ldots ,9\} \)(
see appendix \ref{app_A} for further conventions ) and \( G_{\widehat{a}\widehat{b}\widehat{c}}=\epsilon _{\alpha \beta }V^{\alpha }_{+}F^{\beta }_{\widehat{a}\widehat{b}\widehat{c}} \)
.

It is then a simple matter to solve the equation of motions and to find that

\begin{eqnarray}
\;  & -r_{1}=r_{2}=\mid e\mid ^{2} & \; \label{eq_mot} \\
g_{1}=\frac{2}{3}e & \; \;  & g_{2}=\frac{2}{3}e\rho \; \; (\rho =\pm 1)\label{eq_mot_2} 
\end{eqnarray}
where \( e \) is an arbitrary complex number and to verify that the complex
tensor \( G_{\widehat{a}\widehat{b}\widehat{c}} \) is covariantly constant
in this background, i.e. \( D^{\widehat{a}}G_{\widehat{a}\widehat{b}\widehat{c}}=0 \)
. We notice that the constraint \( g_{1}^{2}=g_{2}^{2} \) is a consequence
of the equation of motion for the scalars.

\subsection{The Killing vectors.}

As a first step we write the supersymmetric transformation rules for the fermions
in this background
\begin{eqnarray}
\delta \lambda  & = & -\frac{i}{8}\Gamma ^{\widehat{a}\widehat{b}\widehat{c}}\widehat{\epsilon }\: G_{\widehat{a}\widehat{b}\widehat{c}}\label{susy_var} \\
\delta \Psi  & = & D\widehat{\epsilon }+\frac{1}{32}\left( 9\: \Gamma ^{\widehat{a}\widehat{b}}V^{\widehat{c}}+\Gamma ^{\widehat{a}\widehat{b}\widehat{c}\widehat{d}}V_{\widehat{d}}\right) \widehat{\epsilon _{c}}\: G_{\widehat{a}\widehat{b}\widehat{c}}
\end{eqnarray}
where \( \widehat{\epsilon }_{c}=\widehat{C}\: \widehat{\epsilon }^{T} \) is
the charge conjugate of the infinitesimal parameters \( \widehat{\epsilon } \)
and all the field strengths are restricted to the spacetime.

Since the gravitino transformation rule involves both the parameters \( \widehat{\epsilon } \)
and its charge conjugate \( \widehat{\epsilon _{c}} \) the only way to set
to zero \( \delta \Psi  \) is to use a parameter \( \widehat{\epsilon } \)
which is a sum of linearly independent Majorana spinors, thus breaking at least
half of the number of local charges, explicitly:
\begin{equation}
\label{ansatz_susy_par}
\widehat{\epsilon }=\left( \begin{array}{c}
0_{16}\\
e^{i\varphi _{N}}\: \epsilon _{N}\otimes \widetilde{\eta }^{N}
\end{array}\right) 
\end{equation}
where \( \epsilon _{N}=C\: \overline{\epsilon }_{N} \) is a set of \( N \)
(\( N \) to be determined) \( AdS_{3} \) Majorana spinors and \( \widetilde{\eta }^{N}=\widetilde{C}\: \widetilde{\overline{\eta }}^{N} \)
is a set of \( N \) \( S_{3}\otimes T^{4} \) Majorana spinors.

The dilatino variation implies that
\begin{equation}
\label{susy_const_1}
1_{2}\otimes \overline{\gamma }_{5}\: \widetilde{\eta }^{N}=\rho \: \widetilde{\eta }^{N}
\end{equation}
where the sign \( \rho  \) is the same as the one which enters eq. (\ref{eq_mot_2}).

The gravitinos variation yields more constraints and suggests how to express
the gravitino (\ref{ansatz_Psi}), when thought as a field on the superspace,
using the \( AdS_{3|4}\otimes S_{3}\otimes T^{4} \) generalized potentials
. Explicitly we get (we denote \( \Psi _{-} \)the 16 lower components of the
antiWeyl gravitino \( \Psi  \))\footnote{
We also define \( D_{AdS_{3}}\: \epsilon =\left( d-\frac{1}{4}\omega ^{ab}\gamma _{ab}\right) \epsilon  \)
and \( D_{S_{3}\otimes T^{4}}\: \widetilde{\eta }=\left( d-\frac{1}{4}\omega ^{ij}\widetilde{\gamma }_{ij}\right) \widetilde{\eta } \)
where \( \omega  \) are the classical spin-connections, i.e. without \( \theta  \)
dependence.
}
\begin{eqnarray*}
\delta \Psi _{_{-}} & = & e^{i\varphi _{N}}\: \left( D_{AdS_{3}}\: \epsilon _{N}+i\frac{e}{2}e^{-2i\varphi _{N}}\: \: \gamma _{a}\epsilon _{N}\: V^{a}\right) \otimes \widetilde{\eta }^{N}\\
 & + & e^{i\varphi _{N}}\: \epsilon _{N}\otimes \left( D_{S_{3}\otimes T^{4}}\: \widetilde{\eta }^{N}-\frac{e}{2}\rho e^{-2i\varphi _{N}}\: \gamma _{i}\otimes 1_{4}\widetilde{\eta }^{N}\: V^{i}\right) 
\end{eqnarray*}
which can be solved as
\begin{eqnarray}
\left( D_{AdS_{3}}\: \epsilon _{N}-i\: s_{N}\: \frac{|e|}{2}\: \gamma _{a}\epsilon _{N}\: E^{a}\right)  & = & 0\label{kill_spin_ads} \\
\left( D_{S_{3}\otimes T^{4}}\: \widetilde{\eta }^{N}+s_{N}\: \frac{|e|}{2}\: \widetilde{\gamma }_{i}\widetilde{\eta }^{N}\: E^{i}\right)  & = & 0\label{kill_spin_s3} 
\end{eqnarray}
where
\begin{equation}
\label{segnante}
-\frac{e}{|e|}\: e^{-2i\varphi _{N}}=s_{N}=\pm 1
\end{equation}
because of the integrability conditions of the two equations (\ref{kill_spin_ads},\ref{kill_spin_s3}).
Notice that in eq.s (\ref{kill_spin_ads},\ref{kill_spin_s3}) we have explicitly
used the fact that the restriction of the zehnbein to the spacetime gives the
classical zehnbein, i.e. \( V^{\widehat{a}}|_{''\theta ''=0}=E^{\widehat{a}} \). 

We will denote a \( S_{3}\otimes T^{4} \) spinor \( \widetilde{\eta }^{N} \)
which satisfies eq. (\ref{kill_spin_s3}) with the substitution \( s_{N}\rightarrow -s \)
as \( \widetilde{\eta }^{N}_{(s)} \) in order to stress the relative sign of
the two terms of the equation (\ref{kill_spin_s3}). 

Before counting the number of real supersymmetric charges surviving in this
background and being able to specify in which range the index \( N \) runs,
we need to understand better how \( \widetilde{\eta }^{N} \)can be decomposed
w.r.t. \( T^{4} \) and \( S_{3} \) spinors. It is not difficult to see that
the Majorana \( \widetilde{\eta }^{N} \)spinors can be decomposed as

\begin{equation}
\label{kill_spin_decomp}
\widetilde{\eta }_{(s)}^{N}=\widetilde{\eta }_{(s)}^{(n,m)}=H^{N}_{pq}\: \eta _{(s)}^{p}\otimes \kappa ^{q}\; \; \; \; \; H^{N}=\sigma _{2}H^{*}_{N}\sigma _{2}
\end{equation}
where the constraint on \( H_{N} \) is due to the Majorana condition which
\( \widetilde{\eta }_{(s)}^{N} \) have to satisfy and to the fact that 

\begin{itemize}
\item \( \kappa ^{q} \) (\( q=1,2 \) ) are a couple of constant chiral symplectic-Majorana
spinors on the 4-torus 
\begin{equation}
\label{kill_spin_t4}
\overline{\gamma }_{5}\: \kappa ^{q}=\rho \: \kappa ^{q}\; \; \; \; \kappa _{c\, q}\equiv \overline{C}\overline{\kappa }_{q}^{T}=\epsilon _{qp}\: \kappa ^{p}\; \; \; \; \overline{\kappa }_{p}\kappa ^{q}=\delta ^{q}_{p}
\end{equation}
with \( \epsilon _{12}=1 \) and 
\item \( \eta ^{p}_{(s)} \) (\( p=1,2 \)) are (for any \( s \)) a couple of symplectic-Majorana
spinors on the 3-sphere
\begin{equation}
\label{simpl-maj-eta}
\eta _{c\, p}\equiv C\: \overline{\eta }_{p}^{T}=\epsilon _{pq}\: \eta ^{q}\; \; \; \; \overline{\eta }_{p}\eta ^{q}=\delta ^{q}_{p}
\end{equation}
which satisfy the killing equation on the 3-sphere
\begin{equation}
\label{kill_spin_s3_pure}
D_{S_{3}}\eta ^{p}_{(s)}-s\frac{|e|}{2}\rho \: \gamma _{i}\eta ^{p}_{(s)}\: E^{i}=0
\end{equation}

\end{itemize}
It is now obvious that both \( p \) and \( q \) in (\ref{kill_spin_decomp})
can take two values and that the index \( N=(n,m) \) can take \( 4 \) values
as we can immediately find 4 matrices which satisfy the second equation in eq.
(\ref{kill_spin_decomp}) \( H \): \( \left\{ 1,\: i\sigma _{1},\: i\sigma _{2},\: i\sigma _{3}\right\}  \)
, hence the number of real supersymmetric charges is \( 2\cdot 2\cdot 4=16 \)
where the first factor \( 2 \) accounts for the two possible values of \( s_{N} \)
(\ref{segnante}) and the second factor is the number of real components of
\( \epsilon _{N} \) .

\subsection{The type IIB fields as fields living on the superspace \protect\( AdS_{3|4}\otimes S_{3}\otimes T^{4}\protect \).}

We are now ready to write our ansatz on how to express the supergravity fields
leaving on superspace, to this purpose we use 

\begin{itemize}
\item the fields of the super anti de Sitter in 3D and 4 complex charges \( AdS_{3|4}=\frac{SU(1,1|2,0)\otimes SU(1,1|0,2)}{SO(1,2)\otimes O(4)} \)
\( \left\{ \omega ^{(\pm )a}(x,\Theta ^{(\pm )}),\: A^{(\pm )n}_{.m}(x,\Theta ^{(\pm )}),\: \chi ^{(\pm )}_{n}(x,\Theta ^{(\pm )})\right\}  \)
(see appendices \ref{app_B} and \ref{app_C} for a complete explanation of
the notations and derivation of these results ) which satisfy
\begin{eqnarray}
d\omega ^{(\pm )a}+\frac{1}{2}\epsilon ^{a}_{.\, bc}\: \omega ^{(\pm )b}\omega ^{(\pm )c}-is^{(\pm )}\: \overline{\chi }^{(\pm )n}\gamma ^{a}\chi _{n}^{(\pm )} & = & 0\nonumber \label{Maure-Cartan} \\
\:  & \:  & \: \label{MC-omega} \\
dA^{(\pm )n}_{.m}+A^{(\pm )n}_{.\, p}\: A^{(\pm )p}_{.\, m}-s^{(\pm )}\: \left( \overline{\chi }^{(\pm )n}\chi ^{(\pm )}_{m}-\frac{1}{2}\delta ^{n}_{m}\: \overline{\chi }^{(\pm )p}\chi ^{(\pm )}_{p}\right)  & = & 0\nonumber \\
\:  & \:  & \: \label{MC-A} \\
dB^{(\pm )}-s^{(\pm )}\frac{1}{2}\: \overline{\chi }^{(\pm )p}\chi ^{(\pm )}_{p} & = & 0\nonumber \\
\:  & \:  & \: \label{MC-B} \\
d\chi ^{(\pm )}_{n}-\frac{i}{2}\omega ^{(\pm )a}\: \gamma _{a}\chi ^{(\pm )}_{n}+\chi ^{(\pm )}_{m}\: A^{(\pm )m}_{.n} & = & 0\nonumber \label{Maurer-Cartan-fine} \\
\:  & \:  & \: \label{MC-chi} 
\end{eqnarray}
and depend on \( x \) and \( \Theta ^{(\pm )} \) which are respectively the
\( AdS_{3} \) coordinates and a set of 4 complex Grassman coordinates which
transform as two \( AdS_{3} \) complex spinors and \emph{not} as \( AdS_{3}\otimes S_{3}\otimes T^{4} \)
spinors; we notice moreover the presence of the arbitrary sign \( s^{(\pm )} \)
;
\item the classical dreibein and spin-connection of \( S_{3} \) (\( E^{i}(y) \)
, \( \omega ^{ij}(y) \) ) which satisfy 
\begin{eqnarray}
dE^{i}-\varpi ^{i}_{.j}E^{j} & = & 0\label{geometria-s3} \\
d\omega ^{ij}-\omega ^{ik}\omega _{k}^{.j} & = & |e|^{2}E^{i}E^{j}\label{geometria-s3-fine} 
\end{eqnarray}
 and depend on the \( S_{3} \) coordinates \( y \) ;
\item the classical vierbein of \( T^{4} \) \( E^{r}=dz^{r} \) where \( z^{r} \)
are the \( T^{4} \) coordinates. 
\end{itemize}
When we take in account the form of the killing spinors (\ref{ansatz_susy_par})
the ansatz for the gravitino reads:

\begin{eqnarray}
\Psi  & = & \kappa _{3}\: \left( \begin{array}{c}
0\\
e^{i\varphi _{(+)}}\chi ^{(+)}_{(n,m)}(x,\Theta )\otimes \widetilde{\eta }^{(n,m)}_{(s)}+e^{i\varphi _{(-)}}\chi ^{(-)}_{(n,m)}(x,\Theta )\otimes \widetilde{\eta }^{(n,m)}_{(t)}
\end{array}\right) =\nonumber \\
 & = & \kappa _{3}\: \left( \begin{array}{c}
0\\
e^{i\varphi _{(+)}}\left( \chi ^{(+)}_{n}(x,\Theta )\otimes \widetilde{\eta }^{n}_{(s)c}+\chi ^{(+)n}_{c}(x,\Theta )\otimes \widetilde{\eta }_{(s)n}\right) 
\end{array}\right) +\nonumber \\
 &  & +\kappa _{3}\: \left( \begin{array}{c}
0\\
e^{i\varphi _{(-)}}\left( \chi ^{(-)}_{n}(x,\Theta )\otimes \widetilde{\eta }^{n}_{(t)c}+\chi ^{(-)n}_{c}(x,\Theta )\otimes \widetilde{\eta }_{(t)n}\right) 
\end{array}\right) \label{ansatz_Psi} 
\end{eqnarray}
where we have defined implicitly the Majorana component of \( \chi _{n} \)
as 
\begin{eqnarray}
\chi _{n} & = & \frac{1}{\sqrt{2}}\left( \chi _{(n,1)}+i\: \chi _{(n,2)}\right) \label{chi-majorana} \\
\chi _{c}^{n} & = & \frac{1}{\sqrt{2}}\left( \chi _{(n,1)}-i\: \chi _{(n,2)}\right) \nonumber 
\end{eqnarray}
and inversely we have defined the complex spinors \( \widetilde{\eta } \) out
of the Majorana ones \( \widetilde{\eta }_{N} \) as
\begin{equation}
\label{eta-complex}
\widetilde{\eta }_{n}=\frac{1}{\sqrt{2}}\left( \widetilde{\eta }_{(n,1)}+i\: \widetilde{\eta }_{(n,2)}\right) =\left( H_{n}\right) _{pq}\: \eta ^{p}\otimes \kappa ^{q}
\end{equation}
and their charge conjugate \( \widetilde{\eta }_{c}^{n} \) . Moreover the Majorana
killing spinors \( \widetilde{\eta }_{(s)N}=\widetilde{\eta }_{(s)(m,l)} \)
(\ref{kill_spin_decomp}) are normalized as 
\begin{equation}
\label{norm-cond-eta}
\widetilde{\overline{\eta }}_{(s)N}\widetilde{\eta }_{(s)M}=\delta _{NM}
\end{equation}

Now that we have a set of complex C-number spinors \( \widetilde{\eta }_{n} \)
with the proper index structure we can write the ansatz for the remaining fields:

\begin{eqnarray}
V^{a} & = & E^{a}(x,\Theta )=\frac{1}{2|e|}\left( \omega ^{(+)a}-\omega ^{(-)a}\right) \label{ansatz_V_1} \\
V^{i} & = & E^{i}(y)+\kappa _{1}^{(+)}\: \widetilde{\overline{\eta }}^{m}_{(s)}(y)\, \widetilde{\gamma }^{i}\, \widetilde{\eta }_{(s)n}(y)\: A^{(+)n}_{\; .\; m}(x,\Theta )\nonumber \\
\:  & \:  & +\kappa _{1}^{(-)}\: \widetilde{\overline{\eta }}_{(t)}^{m}(y)\, \widetilde{\gamma }^{i}\, \widetilde{\eta }_{(t)n}(y)\: A^{(-)n}_{\; .\; m}(x,\Theta )\\
V^{r} & = & dz^{r}\label{ansatz_V_3} \\
\Omega ^{ab} & = & \omega ^{ab}(x,\Theta )=\frac{1}{2}\epsilon ^{abc}\left( \omega _{c}^{(+)}+\omega _{c}^{(-)}\right) \label{ansatz_O_1} \\
\Omega ^{ij} & = & \omega ^{ij}(y)+\kappa _{2}^{(+)}\: \widetilde{\overline{\eta }}_{(s)}^{m}(y)\, \widetilde{\gamma }^{ij}\, \widetilde{\eta }_{(s)n}(y)\: A^{(+)n}_{\; .\; m}(x,\Theta )\nonumber \\
\:  & \:  & +\kappa _{2}^{(-)}\: \widetilde{\overline{\eta }}_{(t)}^{m}(y)\, \widetilde{\gamma }^{ij}\, \widetilde{\eta }_{(t)n}(y)\: A^{(-)n}_{\; .\; m}(x,\Theta )\\
\Omega ^{rs} & = & 0\label{ansatz_O_fine} 
\end{eqnarray}

In order to find the solution we have to fix the coefficients \( \kappa ^{(\pm )}_{1} \),
\( \kappa ^{(\pm )}_{2} \) , \( \kappa _{3} \) (\( \kappa _{3}>0 \) since
its phase is given by \( \varphi _{(+)} \) ) and the phases \( \varphi _{(\pm )} \);
to determine the matrices \( H^{N} \) (\ref{kill_spin_decomp}) and the sign
\( s \) and \( t \); to find to which form of the superalgebra \( SU(1,1|2) \)
the \( \omega ^{(\pm )a}(x,\Theta ) \), \( A^{(\pm )n}_{.m}(x,\Theta ) \),
\( \chi ^{(\pm )}_{n}(x,\Theta ) \) belong, i.e. which sign has \( s^{(\pm )} \)
in the Maurer-Cartan system (\ref{Maure-Cartan}-\ref{Maurer-Cartan-fine}).

Examining the pure gravitational Bianchi identities using these parametrization
(\ref{ansatz_Psi}, \ref{ansatz_V_1}-\ref{ansatz_O_fine}) we get that

\begin{eqnarray}
\kappa _{1}^{(\pm )} & = & \pm \frac{1}{2|e|}\label{k1} \\
\kappa ^{(\pm )}_{2} & = & \frac{1}{2}\label{k2} \\
\kappa _{3} & = & \frac{1}{2\sqrt{|e|}}\label{k3} \\
s=-t & = & -1\\
s^{(\pm )} & = & \pm 1\\
e^{i(\varphi _{(-)}-\varphi _{(+)})} & = & i\label{eifi-fi} \\
H_{1}=\left( \begin{array}{cc}
 & 0\\
-1 & 
\end{array}\right) \;  & \;  & \; H_{2}=\left( \begin{array}{cc}
1 & \\
 & 0
\end{array}\right) \label{H1H2} 
\end{eqnarray}
while the integrability of the \( A \) equation requires

\begin{equation}
\label{e2ifi}
e^{2i\varphi _{(+)}}=-\frac{e}{|e|}
\end{equation}
We do not derive the explicit form of \( A \) here because of its complication
but we will find its explicit expression after we have the \( \kappa  \) fixed
expression for the zehnbein and the gravitini in the next section .

\section{Fixing the \protect\( \kappa \protect \) symmetry using the supersolvable
and the \protect\( \kappa \protect \) gauge fixed expression for the supergravity
background fields.}

In order to find the supersolvable algebras associated with the two forms of
\( SU(1,1|2) \), we look at the solvable algebra associated to \( AdS_{3}=\frac{SO(2,2)}{SO(1,2)}=\frac{SO(1,2)\otimes SO(1,2)}{SO(1,2)} \)
; from (\cite{GruppoToGH}) we see that the generators of \( AdS_{3} \) can
be chosen to be
\[
solv(AdS_{3})=\left\{ M_{23},M_{+0},M_{+1}\right\} \]
where \( M_{\mu \nu } \) are the \( SO(2,2) \) generators and \( M_{+\mu }=\frac{1}{\sqrt{2}}\left( M_{2\mu }+M_{3\mu }\right)  \)
(see appendix \ref{app_B} for further details and more precise notations).
We can now see how these generators can be written in terms of the generators
\( J^{(\pm )}_{c}=\epsilon _{c}^{\, ab}\frac{1}{4}\left( M_{ab}\pm \frac{1}{2}\epsilon _{ab\mu \nu }\: M^{\mu \nu }\right)  \)
of the two \( SO(1,2) \)
\begin{equation}
\label{solv}
solv(AdS_{3})=\left\{ J_{2}^{(+)}-J_{2}^{(-)}\, ,\, -J_{+}^{(+)}\pm J_{-}^{(-)}\right\} 
\end{equation}

This equation (\ref{solv}) can be read in the dual space by saying that the
bosonic part of the supersolvable algebra has to be generated by \( \left\{ \omega ^{(+)2}-\omega ^{(-)2}\, ,\, \omega ^{(+)+}\pm \omega ^{(-)-}\, \right\}  \).
We write down the Maurer-Cartan equations for the two \( SU(1,1|2) \) and we
ask for consistency when we set either \( \omega ^{(+)-} \) or \( \omega ^{(-)+} \).
If we generically denote the two components of the fermionic 1-form \( \chi  \)
as 
\[
\chi =\left( \begin{array}{c}
\zeta \\
\xi 
\end{array}\right) \]
then it easy to find that the \( (+) \) algebra reads

\begin{eqnarray}
d\omega ^{(+)2} & = & 0\label{MC-solv+-1} \\
d\omega ^{(+)+}+\omega ^{(+)+}\omega ^{(+)2}-is^{(+)}\sqrt{2}\xi ^{*n}\xi _{n} & = & 0\label{MC-solv+-2} \\
d\xi ^{(+)}_{n}-\frac{1}{2}\omega ^{(+)2}\xi _{n}^{(+)} & = & 0\label{MC-solv+-3} 
\end{eqnarray}
and the \( (-) \) algebra reads
\begin{eqnarray}
d\omega ^{(-)2} & = & 0\label{MC-solv--1} \\
d\omega ^{(-)-}-\omega ^{(-)-}\omega ^{(-)2}-is^{(-)}\sqrt{2}\zeta ^{*n}\zeta _{n} & = & 0\label{MC-solv--2} \\
d\zeta _{n}^{(-)}+\frac{1}{2}\omega ^{(-)2}\zeta _{n}^{(-)} & = & 0\label{MC-solv--3} 
\end{eqnarray}

From the knowledge of which 1-form make the supersolvable algebra we can deduce
the generic element of the supersolvable algebra (see appendix for \ref{app_C}
notations); for example for the (+) algebra we get 
\[
g^{(+)}=\left( \begin{array}{ccc}
\frac{1}{2}\alpha ^{(+)2} &  & \\
-\frac{i}{\sqrt{2}}\alpha ^{+} & -\frac{1}{2}\alpha ^{(+)2} & \tau _{m}\\
-s^{(+)}\tau ^{*n} &  & 
\end{array}\right) \]
it is now an easy task to derive the explicit dependence on these coordinates
of the (+) 1-forms from the generic group element \( G^{(+)}=\exp \left( g_{2}^{(+)}\right) \exp \left( g_{+}^{(+)}\right) \exp \left( g_{\tau }^{(+)}\right)  \)
:

\begin{eqnarray*}
\omega ^{(+)2} & = & d\alpha ^{(+)2}\\
\omega ^{(+)+} & = & d\alpha ^{+}-\alpha ^{+}d\alpha ^{(+)2}-\frac{i}{\sqrt{2}}s^{(+)}\left( \tau ^{*n}\left( \overrightarrow{d}-\overleftarrow{d}\right) \tau _{n}\right) \\
\xi _{n}^{(+)} & = & d\tau _{n}-\frac{1}{2}\tau _{n}d\alpha ^{(+)2}
\end{eqnarray*}
Similarly for the (-) algebra we get 
\begin{eqnarray*}
\omega ^{(-)2} & = & d\alpha ^{(-)2}\\
\omega ^{(-)-} & = & d\alpha ^{-}+\alpha ^{-}d\alpha ^{(-)2}-\frac{i}{\sqrt{2}}s^{(-)}\left( \theta ^{*n}\left( \overrightarrow{d}-\overleftarrow{d}\right) \theta _{n}\right) \\
\zeta _{n}^{(-)} & = & d\theta _{n}+\frac{1}{2}\theta _{n}d\alpha ^{(-)2}
\end{eqnarray*}

In order to find the \( \kappa  \) gauge fixed expression for the dreibein
and the gravitini we have to relate the coset \( AdS_{3} \) coordinates with
those of the two \( SO(1,2) \): this is done comparing \( b^{a}P_{a}=|e|b^{a}J^{(+)}_{a}-|e|b^{a}J^{(-)}_{a} \)
with \( \alpha ^{(+)a}J^{(+)}_{a}+\alpha ^{(-)a}J^{(-)}_{a} \). In particular
we get \( \alpha ^{(+)2}=-\alpha ^{(-)2}=|e|b^{2} \), \( \alpha ^{(+)+}=|e|b^{+} \)
and \( \alpha ^{(-)-}=-|e|b^{-} \) . Using the relation \( E^{a}=\frac{1}{2|e|}\left( \omega ^{(+)a}-\omega ^{(-)a}\right)  \)
and changing variables as \( r=\exp \left( |e|b^{2}\right)  \) , \( x^{\pm }=\frac{b^{\pm }}{2\, |e|\, r} \),
\( \theta ^{(+)}=\frac{\tau }{|e|\sqrt{2\, r}} \) and \( \theta ^{(-)}=\frac{\theta }{|e|\sqrt{2\, r}} \)
we get the final \( \kappa  \) gauge fixed expression for the dreibein and
the gravitini
\begin{eqnarray}
E^{\pm } & = & |e|r\left( dx^{\pm }\pm \frac{i}{\sqrt{2}}\, s^{(\pm )}\: \theta ^{(\pm )*n}\left( \overrightarrow{d}-\overleftarrow{d}\right) \theta _{(\pm )n}\right) \label{E+-} \\
E^{2} & = & \frac{1}{|e|}\frac{dr}{r}\label{E2} \\
\chi ^{(+)}_{n} & = & |e|\sqrt{2\, r}\left( \begin{array}{c}
0\\
d\theta ^{(+)}_{n}
\end{array}\right) \label{chi+} \\
\chi ^{(-)}_{n} & = & |e|\sqrt{2\, r}\left( \begin{array}{c}
d\theta ^{(-)}_{n}\\
0
\end{array}\right) \label{chi-} 
\end{eqnarray}

With these expressions in hand we try to integrate the \( \kappa  \) gauge
fixed version of the complex 3 field strength
\[
G=\epsilon _{\alpha \beta }V^{\alpha }_{+}dA^{\beta }=G_{\widehat{a}\widehat{b}\widehat{c}}V^{\widehat{a}}V^{\widehat{b}}V^{\widehat{c}}+2i\overline{\Psi }_{c}\Gamma _{\widehat{a}}\Psi V^{\widehat{a}}\]
 (\( V^{\alpha }_{\pm } \) are constant). This is approach is possible because
the operation of pull-back on the surface which fixes the gauge of the \( \kappa  \)
symmetry commutes with the external derivatives.

We notice immediately that there is a part which is closed but not exact as
it is easy to see from the fact that one of the addends is the \( S_{3} \)
volume form . If we use projective coordinates for the 3-sphere \footnote{
Our conventions for the \( S_{3} \) coset manifold with the Killing induced
metric, i.e. negative definite, are ( \( y^{2}=\eta _{ij}y^{i}y^{j}=-\delta _{ij}y^{i}y^{j} \)
)

\begin{eqnarray*}
E^{i} & = & \frac{2}{|e|}\frac{dy^{i}}{1-y^{2}}\\
\varpi ^{ij} & = & \frac{4y^{[i}dy^{j]}}{1-y^{2}}
\end{eqnarray*}

}, the final answer is\footnote{
Actually the sign of the second addend of eq . (\ref{A-1}) is not fixed because
there is a possible sign ambiguity in the value of \( e^{i(\varphi _{(+)}+\varphi _{(-)})} \)
which can be fixed as discussed in app. \ref{app_D=3Dsugra}.
}
\begin{eqnarray}
G=d\left( \epsilon _{\alpha \beta }V^{\alpha }_{+}A^{\beta }\right)  & = & \nonumber \\
=\frac{2}{3}e\rho \epsilon _{ijk}E^{i}E^{j}E^{k}+d\left( \frac{e}{2|e|^{3}}\omega ^{(+)+}\omega ^{(-)-}\right.  & - & \left. \frac{1}{|e|^{2}}e^{i(\varphi _{(+)}+\varphi _{(-)})}\overline{\chi }_{N}^{(+)}\chi _{M}^{(-)}\widetilde{\overline{\eta }}_{(s)N}\widetilde{\eta }_{(-s)M}\right) =\nonumber \\
 &  & \label{A-1} \\
=\frac{16e}{3|e|^{3}} & \rho  & \epsilon _{ijk}\frac{dy^{i}\wedge dy^{j}\wedge dy^{k}}{\left( 1-y^{2}\right) ^{3}}+\nonumber \\
+d\left\{ -2e|e|r^{2}\left[ dx^{+}+\frac{i}{\sqrt{2}}\theta ^{(+)*n}\left( \overrightarrow{d}-\overleftarrow{d}\right) \theta _{(+)n}\right] \right.  & \wedge  & \left[ dx^{-}+\frac{i}{\sqrt{2}}\theta ^{(-)*n}\left( \overrightarrow{d}-\overleftarrow{d}\right) \theta _{(-)n}\right] \nonumber \\
+2i\frac{e}{|e|^{2}}r\left[ d\theta ^{(+)*n}d\theta ^{(-)}_{m}\: \widetilde{\overline{\eta }}_{(s)cn}\widetilde{\eta }_{(-s)c}^{m}\right.  & + & d\theta ^{(+)}_{n}d\theta ^{(-)*m}\: \widetilde{\overline{\eta }}_{(s)}^{n}\widetilde{\eta }_{(-s)m}\nonumber \\
-i\, d\theta _{n}^{(+)}d\theta _{m}^{(-)}\, \widetilde{\overline{\eta }}^{n}_{(s)}\widetilde{\eta }_{(-s)c}^{m} & - & \left. \left. i\, d\theta ^{(+)*n}d\theta ^{(-)*m}\, \widetilde{\overline{\eta }}_{(s)cn}\widetilde{\eta }_{(-s)m}\right] \right\} \label{A-2} 
\end{eqnarray}

\section{The action.}

We have now all the necessary ingredients to be able to write down the action
on this background using the general form given in (\cite{GHMNT}) or the specific
action derived in app. \ref{app_action} (after setting \( \phi =-\frac{\pi }{2} \)):

\begin{eqnarray}
S & = & \int _{\Sigma _{2}}d^{2}\xi \, \sqrt{-g}\, g^{\alpha \beta }\frac{1}{2}\times \nonumber \\
\left\{ 2|e|^{2}\: r^{2}\: \left[ \partial _{\alpha }x^{+}+\frac{i}{\sqrt{2}}\theta ^{(+)*n}\left( \overrightarrow{\partial }_{\alpha }-\overleftarrow{\partial }_{\alpha }\right) \theta _{(+)n}\right] \right.  & \,  & \left[ \partial _{\beta }x^{-}+\frac{i}{\sqrt{2}}\theta ^{(-)*n}\left( \overrightarrow{\partial }_{\beta }-\overleftarrow{\partial }_{\beta }\right) \theta _{(-)n}\right] \nonumber \\
-\frac{1}{|e|^{2}}\frac{\partial _{\alpha }r\: \partial _{\beta }r}{r^{2}} & - & \frac{4}{|e|^{2}}\delta _{ij}\left. \frac{\partial _{\alpha }y^{i}\: \partial _{\beta }y^{j}}{(1-y^{2})^{2}}-\delta _{rs\: }\partial _{\alpha }z^{r}\: \partial _{\beta }z^{s}\right\} \nonumber \\
+|e|\: \Im e\: r^{2}\left[ dx^{+}+\frac{i}{\sqrt{2}}\theta ^{(+)*n}\left( \overrightarrow{d}-\overleftarrow{d}\right) \theta _{(+)n}\right]  & \wedge  & \left[ dx^{-}+\frac{i}{\sqrt{2}}\theta ^{(-)*n}\left( \overrightarrow{d}-\overleftarrow{d}\right) \theta _{(-)n}\right] \nonumber \\
-\frac{\Re e}{|e|}r\left[ d\theta ^{(+)*n}\wedge d\theta ^{(-)}_{m}\: \widetilde{\overline{\eta }}_{(s)cn}\widetilde{\eta }_{(-s)c}^{m}\right.  & + & d\theta ^{(+)}_{n}\wedge d\theta ^{(-)*m}\: \widetilde{\overline{\eta }}^{n}_{(s)}\widetilde{\eta }_{(-s)m}\nonumber \\
-id\theta _{n}^{(+)}\wedge d\theta _{m}^{(-)}\: \widetilde{\overline{\eta }}^{n}_{(s)}\widetilde{\eta }_{(-s)c}^{m} & - & \left. id\theta ^{(+)*n}\wedge d\theta ^{(-)*m}\: \widetilde{\overline{\eta }}_{(s)cn}\widetilde{\eta }_{(-s)m}\right] \nonumber \\
-\frac{1}{3}\rho \Im e\: \int _{B_{3}}\frac{8}{|e|^{3}}\epsilon _{ijk}\frac{dy^{i}\wedge dy^{j}\wedge dy^{k}}{\left( 1-y^{2}\right) ^{3}} & \label{The-Action} 
\end{eqnarray}

We notice that this action is written with the unusual metric which is mostly
minus and with the use of the \( AdS_{3|4} \) fermionic coordinates like in
(\cite{Mio_ads}), we could as well have used fermionic coordinates which transform
as spinors for \( SO(1,2)\otimes SO(3)\otimes SO(4) \) like in ( \cite{Kall_ads}).
This action can be rewritten in a more compact form using \( 6D \) pseudo-Majorana
Weyl spinors when \( \Im e=0 \), this is done in app. \ref{action_D6_fermion}.

It is clear from the explicit form of the action that setting \( \Re e=0 \)
is to choose a pure NSNS background while setting \( \Im e=0 \) is equivalent
to choose a pure RR background, but now the WZW term and the chiral structure
disappear too. The same happened in the case treated previously, i.e. \( AdS_{5}\otimes S_{5} \).
It is therefore tempting to argue that pure RR backgrounds are described by
conformal actions without WZWN term. This is a further hint of fact that in
order to understand pure RR backgrounds requires a lot of efforts.

This action should be conformal because in the supercosets approach there can
be constructed two WZ terms (\cite{MatsaevTseytlin}) whose coefficients, in
proper unities, must be integer, therefore they cannot mix and renormalise but
the kinetic term has a relative weight fixed by the background solution hence
it cannot be renormalized too. As a further evidence of this claim we notice
that in the pure NSNS background the relative normalization of the kinetic term
and the WZWN term is right and equal to \( \frac{1}{3} \).

As derived in app. \ref{app_kappa-symm} or as it can be easily derived requiring
the invertibility of the fermionic kinetic operator in the form of the action
given in eq. (\ref{action_6D}) can only describe the states which satisfy the
condition
\begin{equation}
\label{stati_che_si_descrivono}
2r^{2}\left( E^{-}_{0}-\frac{\Im e}{\left| e\right| }E^{-}_{1}\right) \left( E^{+}_{0}+\frac{\Im e}{\left| e\right| }E^{+}_{1}\right) +\left( \frac{\Re e}{|e|}\right) ^{2}\left( -\left( E^{2}_{1}\right) ^{2}-\delta _{ij}E^{i}_{1}E^{j}_{1}\right) \neq 0
\end{equation}
where \( E^{\hat{a}}_{\alpha } \) is the component of the pullback of \( E^{\hat{a}} \)
on the string surface along \( d\xi ^{\alpha } \) .

In particular for the pure NSNS background it happens that the static solution
(take \( \Im e=-|e| \) for definiteness)
\begin{equation}
\label{static_NSNS}
x^{\pm }=\xi ^{\pm }\: \: ,\: \: r=const\: \: ,\: \: \Theta =0
\end{equation}
cannot be described by the action even if it is a solution of the equations
of motion since it does not satisfy eq. (\ref{stati_che_si_descrivono}) nevertheless
the arguments of (\cite{PST}) do not apply since (\ref{static_NSNS}) is not
a viable solution for a fundamental IIB string which has to be closed, i.e.
\( x^{\pm } \) have to be periodic in \( \sigma  \).

\textbf{Acknowledgements}

It is a pleasure to thank M.L. Frau, A. Lerda, S. Penati, R. Russo and S. Sciuto
for discussions.

\appendix

\section{Conventions.\label{app_A}}

We will use the following conventions:

\begin{itemize}
\item Indices: \( \widehat{a},\widehat{b},\ldots \in \{0,\ldots ,9\} \) , \( a,b,\ldots \in \{0,1,2\} \),
\( u,v,\ldots \in \{3,\ldots ,9\} \), \( i,j,\ldots \in \{3,4,5\} \), \( r,s,\ldots \in \{6,\ldots ,9\} \);
WS indices \( \alpha ,\beta ,\ldots \in \left\{ 0,1\right\}  \)
\item Epsilon: \( \epsilon _{0\ldots 9}=\epsilon _{012}=\epsilon _{345}=1 \); WS
\( \epsilon _{01}=1 \)
\item metric: \( \eta _{\widehat{a}\widehat{b}}=diag(+1,-1,\ldots -1) \), \( \eta _{ab}=diag(+1,-1,-1) \)
, \( \eta _{ij}=diag(-1,-1,-1) \) , \( \eta _{rs}=diag(-1,-1,-1,-1) \); WS
\( \eta _{\alpha \beta }=diag(1,-1) \)
\item lightcone coordinates for \( AdS_{3} \) : \( x^{\pm }=\frac{1}{\sqrt{2}}\left( x^{0}\pm x^{1}\right)  \),
\( \epsilon _{+-2}=-1 \) 
\item 1+9D gamma matrices
\begin{eqnarray*}
\{\Gamma _{\widehat{a}},\Gamma _{\widehat{b}}\} & = & 2\eta _{\widehat{a}\widehat{b}}\\
\Gamma _{11} & = & \left( \begin{array}{cc}
1_{16} & \\
 & -1_{16}
\end{array}\right) \\
\Gamma _{\widehat{a}}^{T}=-\widehat{C}^{-1}\: \Gamma _{\widehat{a}}\: \widehat{C}\: \: \:  & \widehat{C}^{T}=-\widehat{C}\: \: \:  & \widehat{C}^{\dagger }=\widehat{C}
\end{eqnarray*}
 
\item 1+2D (\( AdS_{3}) \) gamma matrices
\begin{eqnarray*}
\{\gamma _{a},\gamma _{b}\} & = & 2\eta _{ab}\\
\gamma _{0}\gamma _{1}\gamma _{2} & = & i\: 1_{2}\\
\gamma _{a}^{T}=-C^{-1}\: \gamma _{a}\: C\: \: \:  & C^{T}=-C\: \: \:  & C^{\dagger }=C
\end{eqnarray*}
explicitly we have
\[
\gamma ^{0}=\sigma _{1}=\left( \begin{array}{cc}
 & 1\\
1 & 
\end{array}\right) \: \: \gamma ^{1}=i\sigma _{2}=\left( \begin{array}{cc}
 & 1\\
-1 & 
\end{array}\right) \: \: \gamma ^{2}=-i\sigma _{3}=\left( \begin{array}{cc}
-i & \\
 & i
\end{array}\right) \: \: C=\sigma _{2}\]

\item 0+7D (\( S_{3}\otimes T^{4} \)) gamma matrices
\begin{eqnarray*}
\{\widetilde{\gamma }_{u},\widetilde{\gamma }_{v}\} & = & 2\eta _{uv}=-2\delta _{uv}\\
\widetilde{\gamma }_{3}\ldots \widetilde{\gamma }_{8} & = & 1_{8}\\
\widetilde{\gamma }_{u}^{T}=-\widetilde{C}^{-1}\: \widetilde{\gamma }_{u}\: \widetilde{C}\: \: \:  & \widetilde{C}^{T}=+\widetilde{C}\: \: \:  & \widetilde{C}^{\dagger }=\widetilde{C}
\end{eqnarray*}
Moreover given a \( S_{3} \) spinor \( \eta  \) we use the notation \( \overline{\eta }=\eta ^{\dagger } \)
.
\item 0+3D (\( S_{3} \)) gamma matrices
\begin{eqnarray*}
\{\gamma _{i},\gamma _{j}\} & = & 2\eta _{ij}\\
\gamma _{3}\gamma _{4}\gamma _{5} & = & 1_{2}\\
\gamma _{i}^{T}=-C^{-1}\: \gamma _{i}\: C\: \: \:  & C^{T}=-C\: \: \:  & C^{\dagger }=C
\end{eqnarray*}

\item 0+4D (\( T^{4} \)) gamma matrices
\begin{eqnarray*}
\{\gamma _{r},\gamma _{s}\} & = & 2\eta _{rs}\\
\gamma _{6}\gamma _{7}\gamma _{8}\gamma _{9} & = & \overline{\gamma }_{5}\\
\overline{\gamma }_{5}^{2}=1_{4} &  & \overline{\gamma }_{5}^{\dagger }=\overline{\gamma }_{5}\\
\gamma _{r}^{T}=-\overline{C}^{-1}\: \gamma _{r}\: \overline{C} & \: \: \:  & \overline{\gamma }_{5}^{T}=+\overline{C}^{-1}\: \overline{\gamma }_{5}\: \overline{C}\\
\overline{C}^{T}=-\overline{C} & \: \: \:  & \overline{C}^{\dagger }=\overline{C}
\end{eqnarray*}

\item 1+9D gamma matrices expressed using 1+2D and 0+7D gamma matrices
\begin{eqnarray*}
\Gamma _{a} & = & \gamma _{a}\otimes 1_{8}\otimes \sigma _{1}\\
\Gamma _{u} & = & 1\otimes \widetilde{\gamma }_{u}\otimes (-\sigma _{2})\\
\widehat{C} & = & C\otimes \widetilde{C}\otimes \sigma _{1}
\end{eqnarray*}
 
\item 0+7D gamma matrices expressed using 0+3D and 0+4D gamma matrices
\begin{eqnarray*}
\widetilde{\gamma }_{i} & = & \gamma _{i}\otimes \overline{\gamma }_{5}\\
\widetilde{\gamma }_{r} & = & 1_{2}\otimes \gamma _{r}\\
\widetilde{C} & = & C\otimes \overline{C}
\end{eqnarray*}

\item 1+9D gamma matrices expressed using 1+2D , 0+3D and 0+4D gamma matrices
\begin{eqnarray*}
\Gamma _{a} & = & \gamma _{a}\otimes (1_{2}\otimes 1_{4})\otimes \sigma _{1}\\
\Gamma _{i} & = & 1\otimes (\gamma _{i}\otimes \overline{\gamma }_{5})\otimes (-\sigma _{2})\\
\Gamma _{r} & = & 1\otimes (1_{2}\otimes \gamma _{r})\otimes (-\sigma _{2})\\
\widehat{C} & = & C\otimes (C\otimes \overline{C})\otimes \sigma _{1}
\end{eqnarray*}

\end{itemize}

\section{Decomposition \protect\( so(2,2)=so(1,3)\oplus so(1,3)\protect \) and \protect\( AdS_{3}\protect \)
.\label{app_B}}

We normalize the \( so(2,2) \) generators as
\[
[\: M_{\mu \nu }\: ,\: M_{\rho \sigma }\: ]=-4\: \eta _{\begin{array}{cc}
[\mu  & \\
 & [\rho 
\end{array}}M_{\begin{array}{cc}
\nu ] & \\
\bullet  & \sigma ]
\end{array}}\]
where \( \eta =diag(+,-,-+) \) with \( \mu ,\nu \in \{0,1,2,3\} \) . Notice
that these indices have a priori nothing to do with the spatial indices of \( AdS_{3} \),
but in view of the identification of \( AdS_{3} \) as a coset \( \frac{SO(2,2)}{SO(1,2)} \)
we can write
\begin{eqnarray*}
P_{a} & = & |e|\: M_{a3}\\
L_{ab} & = & M_{ab}
\end{eqnarray*}
where \( L \) are the Lorentz generators, \( P \) are the translations , \( |e|^{2} \)
is the curvature and the indices \( a,b,\ldots \in \{0,1,2\} \) are the tangential
indices to \( AdS_{3} \).

We can now define the linear combinations
\[
J^{(\pm )}_{ab}=\epsilon _{ab.}^{\cdot \cdot c}\: J^{(\pm )}_{c}=\frac{1}{2}\left( M_{ab}\pm \frac{1}{2}\epsilon _{ab\mu \nu }\: M^{\mu \nu }\right) \]
where \( \epsilon _{0123}=+1 \) and easily verify that \( J^{(\pm )}_{c} \)
verify a \( so(1,2) \) algebra given by
\[
[\: J_{a\: },\: J_{b}\: ]=-\epsilon _{abc}\: J^{c}\]
with \( \epsilon _{012}=+1 \) .Since this is a change of basis we can express
the \( M \) s through the \( J \) s as
\begin{eqnarray*}
M_{ab} & = & J^{(+)}_{ab}+J^{(-)}_{ab}=\epsilon _{abc}\: \left( J^{(+)c}+J^{(-)c}\right) \\
M_{a3} & = & \frac{1}{2}\epsilon _{abc}\: \left( J^{(+)bc}-J^{(-)bc}\right) =J^{(+)}_{a}-J^{(-)}_{a}
\end{eqnarray*}
 This change of basis reflects on the dual space in the following way
\begin{eqnarray*}
\Omega ^{ab} & = & \frac{1}{2}\left( \omega ^{(+)ab}+\omega ^{(-)ab}\right) =\frac{1}{2}\epsilon ^{ab}_{\cdot \cdot c}\: \left( \omega ^{(+)c}+\omega ^{(-)c}\right) \\
\Omega ^{a3}=|e|\: E^{a} & = & \frac{1}{4}\epsilon ^{a}_{\cdot bc}\: \left( \omega ^{(+)bc}-\omega ^{(-)bc}\right) =\frac{1}{2}\left( \omega ^{(+)a}-\omega ^{(-)a}\right) 
\end{eqnarray*}
This can be easily obtained comparing the decompositions of a generic \( so(2,2) \)
valued 1-form \( \mu =-\frac{1}{2}\Omega ^{\mu \nu }M_{\mu \nu }=-\frac{1}{2}\Omega ^{ab}L_{ab}-E^{a}P_{a} \)
(the coefficient \( -\frac{1}{2} \) is chosen in order to get the usual expression
for the curvature when computing the Maurer-Cartan) with the direct sum of two
generic \( so(1,3) \) valued 1-forms \( \mu =-\frac{1}{2}\omega ^{(+)ab}J^{(+)}_{ab}-\frac{1}{2}\omega ^{(-)ab}J^{(-)}_{ab}=-\frac{1}{2}\omega ^{(+)a}J^{(+)}_{a}-\frac{1}{2}\omega ^{(-)a}J^{(-)}_{a} \).

\section{The \protect\( SU(1,1|2,0)\protect \) and \protect\( SU(1,1|0,2)\protect \)
Maurer-Cartan.\label{app_C}}

We define the two algebras as the set of supermatrices \( g \) satisfying
\[
g^{\dagger }G+Gg=0\; \; \; \; \; G=\left( \begin{array}{cc}
\gamma _{0} & \\
 & s\: 1_{2}
\end{array}\right) \]
where \( s=1 \) for \( SU(1,1|2,0) \) and \( s=-1 \) for \( SU(1,1|0,2) \).We
can now easily write down the most general element as
\[
g=\left( \begin{array}{cc}
-\alpha ^{a}\frac{i}{2}\gamma _{a} & \Theta _{m}\\
-s\: \overline{\Theta }^{n} & \alpha ^{n}_{.\, m}
\end{array}\right) +\beta \: 1_{4}\; \; \; \; \; \alpha ^{a*}=\alpha ^{a}\; ,\; \left( \alpha ^{n}_{.\, m}\right) ^{*}=-\alpha ^{m}_{.\, n}\; \alpha ^{m}_{m}=0,\; \beta ^{*}=-\beta \]
where \( \alpha ^{a} \) (\( a\in \{0,1,2\} \)), \( \alpha ^{n}_{m} \) (\( m,n\in \{1,2\} \)
) and \( \beta  \) are bosonic while \( \Theta _{n}=\left( \begin{array}{c}
\theta _{n}\\
\tau _{n}
\end{array}\right)  \) are 2 3D spinors, i.e. \( \theta  \) and \( \tau  \) are Grassman variables
and \( \overline{\Theta }^{n}=\Theta ^{\dagger n}\gamma _{0} \) .

Given the previous expression for the generic element of the algebra we could
compute the left invariant forms which satisfy the Maurer-Cartan equation in
the standard way as \( \mu (\alpha ^{a},\alpha ^{n}_{m},\beta ,\Theta _{n})=\exp (-g)\, d\exp (g) \)
but, since we are interested only in the expression for the Maurer-Cartan equation,
we need only the expression for a generic (super)Lie algebra valued 1-form

\[
\mu =\left( \begin{array}{cc}
-\omega ^{a}\frac{i}{2}\gamma _{a} & \chi _{m}\\
-s\: \overline{\chi }^{n} & A^{n}_{.\, m}
\end{array}\right) +B\: 1_{4}\; \; \; \; A^{\dagger }=-A\; ,\; B^{*}=-B\]
where the 1-forms \( \omega  \), \( B \) and \( A \) are bosonic while the
1-form \( \chi  \) is a fermionic spinorial 1-form and we have defined \( \overline{\chi }^{n}=\chi ^{\dagger n}\gamma _{0} \)
. From the usual expression for the the Maurer-Cartan equation \( d\mu +\mu \mu =0 \)
we deduce the Maurer-Cartan equations as

\begin{eqnarray*}
d\omega ^{a}+\frac{1}{2}\epsilon ^{a}_{.\, bc}\: \omega ^{b}\omega ^{c}-is\: \overline{\chi }^{n}\gamma ^{a}\chi _{n} & = & 0\\
dA^{n}_{.m}+A^{n}_{.\, p}\: A^{p}_{.\, m}-s\left( \overline{\chi }^{n}\chi _{m}-\frac{1}{2}\delta ^{n}_{m}\overline{\chi }^{p}\chi _{p}\right)  & = & 0\\
dB-s\frac{1}{2}\overline{\chi }^{p}\chi _{p} & = & 0\\
d\chi _{n}-\frac{i}{2}\omega ^{a}\gamma _{a}\chi _{n}+\chi _{m}A^{m}_{.n} & = & 0
\end{eqnarray*}
The first and last equations can be rewritten using \( \omega ^{ab}=\epsilon ^{abc}\omega _{c} \)
in a more standard way as

\begin{eqnarray*}
d\omega ^{ab}-\omega ^{a}_{.\, c}\omega ^{cb}-s\: \overline{\chi }^{n}\gamma ^{ab}\chi _{n} & = & 0\\
d\chi _{n}-\frac{1}{4}\omega ^{ab}\gamma _{ab}\chi _{n}+\chi _{m}A^{m}_{.n} & = & 0
\end{eqnarray*}

\section{Derivation of the supergravitational background.\label{app_D=3Dsugra}}

As a first step we record here two proprerties which turn out very important
in the following

\begin{eqnarray}
\overline{\eta }^{1}\gamma ^{i}\eta _{1}+\overline{\eta }^{2}\gamma ^{i}\eta _{2} & = & 0\label{simm-1} \\
A^{1}_{1}+A^{2}_{2} & = & 0\label{simm-2} 
\end{eqnarray}
the first one is a consequence of the symplectic-Majorana condition (\ref{simpl-maj-eta})
satisfied by \( \eta  \) while the second equation is a consequence of the
definition of \( A \) as a \( SU(2) \) gauge connection.

The projection of the torsion equation on \( AdS_{3} \) yields

\begin{eqnarray}
T^{a} & = & \frac{1}{2|e|}\left[ \left( d\omega ^{(+)a}+\frac{1}{2}\epsilon ^{abc}\omega ^{(+)}_{b}\omega ^{(+)}_{c}\right) -\left( d\omega ^{(-)a}+\frac{1}{2}\epsilon ^{abc}\omega ^{(-)}_{b}\omega ^{(-)}_{c}\right) \right] \nonumber \\
 &  & -i\overline{\Psi }\Gamma ^{a}\Psi =\nonumber \\
 & = & \frac{1}{2|e|}\left[ is^{(+)}\overline{\chi }^{(+)n}\gamma ^{a}\chi ^{(+)}_{n}-is^{(-)}\overline{\chi }^{(-)n}\gamma ^{a}\chi ^{(-)}_{n}\right] -\nonumber \\
 &  & -2i\kappa _{3}^{2}\left[ \overline{\chi }^{(+)n}\gamma ^{a}\chi ^{(+)}_{n}+\overline{\chi }^{(-)n}\gamma ^{a}\chi ^{(-)}_{n}\right] \label{torsion-ads3} 
\end{eqnarray}
where we have used (\ref{Maure-Cartan}) to eliminate \( \omega ^{(\pm )} \)
and we have imposed
\begin{equation}
\label{e2fi-fi}
e^{2i(\varphi _{(-)}-\varphi _{(+)})}=-1
\end{equation}
in order to eliminate the mixed gravitini current \( \overline{\chi }^{(+)}\chi ^{(-)} \)in
the current \( \overline{\Psi }\Gamma ^{a}\Psi  \). We notice that eq. (\ref{eifi-fi})
is not the only solution of eq. (\ref{e2fi-fi}), the other solution with the
r.h.s. of eq. (\ref{eifi-fi}) equal to \( -i \) does not yield a different
solution since the sign can be reabsorbed in the field redefinition \( \chi ^{(-)}\rightarrow -\chi ^{(-)} \).

From eq. (\ref{torsion-ads3}) we can now easily derive the conditions

\begin{eqnarray}
s^{(\pm )} & = & \pm 1\label{s+-} \\
\kappa _{3} & = & \frac{1}{2\sqrt{|e|}}\label{app-k3} 
\end{eqnarray}
The first condition is quite important because it means that \( SU(1,1|2)^{2} \)
is \emph{not} the proper global symmetry of this background.

We can now turn to exam the projection of the torsion on \( S_{3} \); when
we use 
\begin{eqnarray*}
s & = & -t\\
\kappa ^{(\pm )}_{2} & = & \mp s|e|\kappa ^{(\pm )}_{1}
\end{eqnarray*}
in the expression of the torsion we can eliminate the sectors \( A^{(\pm )}E^{i} \)
and \( A^{(+)}A^{(-)} \) and we are left with
\begin{eqnarray*}
T^{i} & = & \kappa ^{(+)}_{1}\left[ \widetilde{\overline{\eta }}_{(s)}^{m}\, \widetilde{\gamma }^{i}\, \widetilde{\eta }_{(s)n}\: dA^{(+)n}_{\; .\; m}-\kappa _{2}^{(+)}\widetilde{\overline{\eta }}_{(s)}^{m}\, \widetilde{\gamma }^{ij}\, \widetilde{\eta }_{(s)n}\: \widetilde{\overline{\eta }}_{(s)}^{p}\, \widetilde{\gamma }_{j}\, \widetilde{\eta }_{(s)q}\: A^{(+)n}_{\; .\; m}\: A^{(+)q}_{\; .\; p}\right] +\\
 &  & +\kappa ^{(-)}_{1}\left[ \widetilde{\overline{\eta }}_{(t)}^{m}\, \widetilde{\gamma }^{i}\, \widetilde{\eta }_{(t)n}\: dA^{(-)n}_{\; .\; m}-\kappa _{2}^{(-)}\widetilde{\overline{\eta }}_{(t)}^{m}\, \widetilde{\gamma }^{ij}\, \widetilde{\eta }_{(t)n}\: \widetilde{\overline{\eta }}_{(t)}^{p}\, \widetilde{\gamma }_{j}\, \widetilde{\eta }_{(t)q}\: A^{(-)n}_{\; .\; m}\: A^{(-)q}_{\; .\; p}\right] +\\
 &  & -i\overline{\Psi }\Gamma ^{i}\Psi 
\end{eqnarray*}
now in order to be able to use eq. (\ref{MC-A}) to eliminate the \( A \) in
favor of \( \chi  \) we need a Fierz identity like the first one of the two
identities
\begin{eqnarray}
\widetilde{\gamma }^{j}\, \widetilde{\eta }_{(r)n}\: \widetilde{\overline{\eta }}_{(r)}^{p}\, \widetilde{\gamma }_{j}\, \widetilde{\eta }_{(r)q} & = & 2\kappa _{4}\: \delta ^{p}_{n}\: \widetilde{\eta }_{(r)q}+\delta ^{p}_{q}\: (\ldots )\label{fierz-1} \\
\widetilde{\gamma }^{j}\, \widetilde{\eta }_{(r)c}^{n}\: \widetilde{\overline{\eta }}_{(r)}^{p}\, \widetilde{\gamma }_{j}\, \widetilde{\eta }_{(r)q} & = & 2\kappa _{5}\: \delta ^{n}_{q}\: \widetilde{\eta }^{p}_{(r)c}+\delta ^{p}_{q}\: (\ldots )\label{fierz-2} 
\end{eqnarray}
which have to be valid for \( r=s,t \) and where \( (\ldots ) \) stands for
something of which it does not matter to know the actual value (as far as it
is independent of \( p,q \) ) in view of eq. (\ref{simm-2}); we notice moreover
that the second equation (\ref{fierz-2}) turns out to be necessary for verifying
the gravitino Bianchi identity. These equations (\ref{fierz-1},\ref{fierz-2})
can be rewritten in terms of \( H_{n} \) (\ref{eta-complex}) as

\begin{eqnarray*}
H_{q}H^{\dagger p}H_{n} & = & \kappa _{4}\: \delta ^{p}_{n}\: H_{q}+\kappa _{6}\: \delta ^{p}_{q}\: H_{n}\\
H_{q}H^{\dagger p}H^{n}_{c} & = & \kappa _{5}\: \delta ^{n}_{q}\: H_{c}^{p}+\kappa _{7}\: \delta ^{p}_{q}\: H_{c}^{n}
\end{eqnarray*}
where we have filled the \( (\ldots ) \) with the only objects with the right
transformation properties that could be used. We have moreover defined \( H_{c}^{n} \)
to be the matrices entering eq. (\ref{eta-complex}) in place of \( H^{n} \)
when we define \( \eta ^{n}_{c} \) ; explictely \( \sigma _{2}H^{*n}\sigma _{2}=H^{n}_{c} \)
.

Since these equations are cubic and therefore difficult to solve it is better
to try to find further constraints on the matrices \( H_{N} \) (\ref{kill_spin_decomp});
some constraints come from the ansatz that the current \( \overline{\Psi }\Gamma ^{i}\Psi  \)
be proportional to \( \overline{\chi }^{(+)m}\chi ^{(+)}_{n}\: \overline{\eta }_{(s)}^{n}\gamma ^{i}\eta _{(s)m}+\overline{\chi }^{(-)m}\chi ^{(-)}_{n}\: \overline{\eta }_{(t)}^{n}\gamma ^{i}\eta _{(t)m} \)
without currents like \( \overline{\chi }^{n}\chi ^{m}_{c} \) :
\begin{eqnarray}
\left( H^{m}_{c}H^{\dagger }_{cn}-H_{n}H^{\dagger m}\right) _{pq} & = & a(\delta ^{m}_{p}\delta ^{q}_{n}-\epsilon _{np}\epsilon ^{mq})+c^{m}_{n}\delta ^{p}_{q}\label{const-H-1} \\
H_{m}H^{\dagger }_{cn}-H_{n}H^{\dagger }_{cm} & = & c_{nm}1\label{const-H-2} 
\end{eqnarray}
 To these constraints we have to add the normalization conditions (\ref{kill_spin_t4},\ref{simpl-maj-eta},\ref{norm-cond-eta})
which in this language read:

\begin{eqnarray}
tr(H^{\dagger n}H_{m}) & = & \delta ^{n}_{m}\label{norm-eta-1} \\
tr(H^{\dagger n}H^{m}_{c}) & = & 0\label{norm-eta-2} 
\end{eqnarray}

As a consequence of eq.s (\ref{norm-eta-1},\ref{norm-eta-2}) and eq. (\ref{const-H-1})
we deduce that
\begin{eqnarray}
H_{1}H^{\dagger 1} & = & \frac{1}{2}1_{2}-\frac{1}{2}a\sigma _{3}\label{HH-inizio} \\
H_{2}H^{\dagger 2} & = & \frac{1}{2}1_{2}+\frac{1}{2}a\sigma _{3}\\
H_{1}H^{\dagger 2} & = & -a\sigma _{-}=-a(\sigma _{1}-i\sigma _{2})\\
H_{2}H^{\dagger 1} & = & -a\sigma _{+}\label{HH-fine} 
\end{eqnarray}

Using these equations (\ref{HH-inizio}-\ref{HH-fine}) into (\ref{fierz-1},\ref{fierz-2})
we get a over constrained system whose solution is :
\begin{eqnarray*}
a=1\; \; \; \kappa _{4}=1\; \; \; \kappa _{5} & = & -1\; \; \; \kappa _{6}=0\; \; \; \kappa _{7}=1\\
H_{1}=\left( \begin{array}{cc}
0 & 0\\
-h_{2} & h_{1}
\end{array}\right) \; \;  & \;  & \; \; H_{2}=\left( \begin{array}{cc}
h_{2} & -h_{1}\\
0 & 0
\end{array}\right) \\
H^{1}_{c}=\left( \begin{array}{cc}
h_{1}^{*} & h_{2}^{*}\\
0 & 0
\end{array}\right) \; \;  & \;  & \; \; H_{c}^{2}=\left( \begin{array}{cc}
0 & 0\\
h_{1}^{*} & h_{2}^{*}
\end{array}\right) 
\end{eqnarray*}
with \( |h_{1}|^{2}+|h_{2}|^{2}=1 \) . The general solution can be obtained
by a \( SU(2) \) rotation applied to \( (H_{n}) \) thought as a two component
complex vector whose entries are matrices.

We can now perform a \( USp(2) \) rotation on \( \eta  \) and \( \kappa  \)
(which leaves invariant the symplectic-Majorana and normalization conditions)
to bring the previous \( H \) to the ones given in the text (\ref{H1H2}).

\section{The GS superstring action on \protect\( AdS_{3}\otimes S_{3}\otimes T^{4}\protect \)
in first order formalism and its \protect\( \kappa \protect \) symmetry.\label{app_action}}

it is purpose of this appendix to determine the explicit form of the superstring
action propagating on \( AdS_{3}\otimes S_{3}\otimes T^{4} \) and the \( \kappa  \)
symmetry transformation rules for its fields. This can be done in two different
ways either starting in \( D=10 \) and then compactifying or directly in \( D=3 \). 

We start with the former approach writing an ansatz inspired by the knowledge
of the second order action (\cite{GHMNT}) using the notations of (\cite{IIB})
and from the fact that the scalars are constant:
\begin{eqnarray}
S & = & \int _{\Sigma }c_{1}\Pi ^{\widehat{a}}_{\alpha }\: i_{*}V^{\widehat{b}}\wedge e^{\beta }\: \eta _{\widehat{a}\widehat{b}}\: \epsilon _{\alpha \beta }+c_{2}\Pi ^{\widehat{a}}_{\alpha }\Pi ^{\widehat{b}}_{\beta }\: \eta _{\widehat{a}\widehat{b}}\: \eta ^{\alpha \beta }\: \epsilon _{\gamma \delta }\: e^{\gamma }\wedge e^{\delta }\nonumber \\
 &  & +c_{3}V_{+}*A-c_{3}^{*}V_{-}*A\label{S-primo-ordine} 
\end{eqnarray}
where \( V_{\pm }*A=\epsilon _{\alpha \beta }V_{\pm }^{\alpha }A^{\beta } \),
\( \alpha ,\beta ,\ldots \in \left\{ 0,1\right\}  \) are worldsheet indexes
and \( i_{*} \) is the pullback on the string worldsheet \( \Sigma  \) due
to the superimmersion \( i:\: \Sigma \: \rightarrow \: AdS_{3}\otimes S_{3}\otimes T^{4} \).

The relative coefficient of the first two terms of the first line in eq. (\ref{S-primo-ordine})
is determined to be
\begin{equation}
\label{c2}
c_{2}=-\frac{1}{4}c_{1}
\end{equation}
by the request that the \( \Pi _{\alpha }^{\widehat{a}} \) equation of motion
yields
\begin{equation}
\label{eq-mot-Pi}
\frac{\delta S}{\delta \Pi }=0\Longrightarrow i_{*}V^{\widehat{a}}=\Pi ^{\widehat{a}}_{\alpha }e^{\alpha }
\end{equation}

The zweibein equation of motion gives the Virasoro constraints:
\begin{equation}
\label{Virasoro-const}
\frac{\delta S}{\delta e}=0\Longrightarrow \Pi _{\alpha }\bullet \Pi _{\beta }=\frac{1}{2}\eta _{\alpha \beta }\Pi ^{2}
\end{equation}

From the request of the action (\ref{S-primo-ordine}) (in the 1.5 formalism)
to have a \( \kappa  \) symmetry we determine the relative value of the coefficients
of last two terms in eq. (\ref{S-primo-ordine}) with respect the first term
in the first line: 
\begin{equation}
\label{c3}
c_{3}=-\frac{1}{4}e^{i\phi }c_{1}\; \; \; \; \forall \phi 
\end{equation}
and the vector field \( \overrightarrow{\widehat{\epsilon }}=\delta _{\kappa }\Theta \: \overrightarrow{\partial _{\Theta }}+\delta _{\kappa }\Theta _{c}\: \overrightarrow{\partial _{\Theta _{c}}}+\delta _{\kappa }x\: \overrightarrow{\partial _{x}} \)
associated to the \( \kappa  \) symmetry and the variation of the zweibein:
\[
\delta _{\kappa }e^{\alpha }=2ic_{1}\: \left( \overline{\widehat{\kappa }}^{\alpha }\Psi +\overline{\widehat{\kappa }_{c}}^{\alpha }\Psi _{c}\right) \]
The vector field \( \overrightarrow{\widehat{\epsilon }} \) is defined by the
properties 
\begin{eqnarray}
_{\overrightarrow{\widehat{\epsilon }}}\mid \psi  & = & \widehat{\epsilon }=\Pi ^{\widehat{a}}_{\alpha }\: \Gamma _{\widehat{a}}\widehat{\kappa }^{\alpha }\: ,\: \: \: _{\overrightarrow{\epsilon }}\mid V^{\hat{a}}=0\label{kappa-symmetry} \\
\widehat{\kappa }_{c\alpha } & = & ie^{i\phi }\epsilon _{\alpha \beta }\: \widehat{\kappa }_{\beta }\label{vincolo-su-kappa} 
\end{eqnarray}
and its action by mean of the Lie derivative gives the supersymmetric variations
(up to local Lorentz and gauge transformations) of the other fields 
\begin{eqnarray}
\delta _{\kappa }V^{\widehat{a}} & = & -i\: \left( \overline{\Psi }\Gamma ^{\hat{a}}\widehat{\epsilon }-\overline{\Psi _{c}}\Gamma ^{\hat{a}}\hat{\epsilon }_{c}\right) \label{kappa-var-1} \\
\delta _{\kappa }(V_{+}*A) & = & -4i\: \overline{\Psi }\: \Gamma ^{\hat{a}}\widehat{\epsilon }V_{\hat{a}}\label{kappa-var-5} 
\end{eqnarray}
We have now to rewrite these expressions using the fields which we used in the
rest of this paper and are proper for treating the background; to this purpose
inspired by the expression for the gravitini eq. (\ref{ansatz_Psi}) we write

\begin{eqnarray}
\widehat{\epsilon }=\kappa _{3}\: \left( \begin{array}{c}
0\\
e^{i\varphi _{(+)}}\epsilon ^{(+)}_{(n,m)}(x,\Theta )\otimes \widetilde{\eta }^{(n,m)}_{(s)}+e^{i\varphi _{(-)}}\epsilon ^{(-)}_{(n,m)}(x,\Theta )\otimes \widetilde{\eta }^{(n,m)}_{(t)}
\end{array}\right)  &  & \label{ansatz-epsilon} \\
\hat{\kappa }=\kappa _{3}\: \left( \begin{array}{c}
e^{i\varphi _{(+)}}\kappa ^{(+)}_{(n,m)}(x,\Theta )\otimes \widetilde{\eta }^{(n,m)}_{(s)}+e^{i\varphi _{(-)}}\kappa ^{(-)}_{(n,m)}(x,\Theta )\otimes \widetilde{\eta }^{(n,m)}_{(t)}\\
0
\end{array}\right)  &  & \label{ansatz-kappa} 
\end{eqnarray}
where we impose Majorana conditions on \( \epsilon _{N}^{(\pm )} \)and antiMajorana
on \( \kappa _{N}^{(\pm )} \) because of the presence of a \( \Gamma  \) in
eq. (\ref{kappa-symmetry}). In order to check the validity of this ansatz we
insert these expressions in the susy transformation rules (\ref{kappa-var-1}-\ref{kappa-var-5})
and we get the same susy transformation rules for the \( AdS_{3}\otimes S_{3}\otimes T^{4} \)
fields as those which can be derived directly from the curvatures (\ref{Maure-Cartan}-\ref{Maurer-Cartan-fine})
and from the expressions (\ref{ansatz_V_1}-\ref{ansatz_V_3}):
\begin{eqnarray}
\delta V^{a} & = & -\frac{i}{2\left| e\right| }\: \left( \overline{\chi }_{N}^{(+)}\gamma ^{a}\epsilon ^{(+)}_{N}+\overline{\chi }_{N}^{(-)}\gamma ^{a}\epsilon ^{(-)}_{N}\right) \label{susy-var-3D-1} \\
\delta V^{i} & = & \frac{1}{2\left| e\right| }\: \left( \overline{\chi }_{N}^{(+)}\epsilon ^{(+)}_{M}K^{i}_{(s)MN}+\overline{\chi }_{N}^{(-)}\epsilon ^{(-)}_{M}K^{i}_{(t)MN}\right) \\
\delta V^{r} & = & 0\\
\delta (V_{+}*A) & = & -\frac{i}{\left| e\right| }\: \left( e^{2i\phi _{(+)}}\overline{\chi }_{N}^{(+)}\gamma ^{a}\epsilon ^{(+)}_{N}+e^{2i\phi _{(-)}}\overline{\chi }_{N}^{(-)}\gamma ^{a}\epsilon ^{(-)}_{N}\right) \: V_{a}\nonumber \label{kappa-var-3D-3} \\
 &  & -\frac{i}{2\left| e\right| }e^{i\left( \phi _{(+)}+\phi _{(-)}\right) }\: \left( \overline{\chi }_{N}^{(+)}\gamma ^{a}\epsilon ^{(-)}_{M}+\overline{\chi }_{N}^{(-)}\gamma ^{a}\epsilon ^{(+)}_{M}\right) \: J_{NM}\: V_{a}\nonumber \\
 &  & +\frac{1}{\left| e\right| }\: \left( e^{2i\phi _{(+)}}\overline{\chi }_{N}^{(+)}\epsilon ^{(+)}_{M}K^{i}_{(s)MN}+e^{2i\phi _{(-)}}\overline{\chi }_{N}^{(-)}\epsilon ^{(-)}_{M}K^{i}_{(t)MN}\right) \: V_{i}\nonumber \\
 &  & +\frac{1}{\left| e\right| }e^{i\left( \phi _{(+)}+\phi _{(-)}\right) }\: \left( \overline{\chi }_{N}^{(+)}\epsilon ^{(-)}_{M}+\overline{\chi }_{N}^{(-)}\epsilon ^{(+)}_{M}\right) \: J^{i}_{NM}\: V_{i}\label{susy-var-3D-4} 
\end{eqnarray}
where we have defined the following quantities:
\begin{eqnarray*}
J_{NM}=\widetilde{\overline{\eta }}^{N}_{(s)}\widetilde{\eta }^{N}_{(t)} & \: \: \:  & J^{i}_{NM}=\widetilde{\overline{\eta }}^{N}_{(s)}\widetilde{\gamma }^{i}\widetilde{\eta }^{N}_{(t)}\\
K^{i}_{(s)MN}=\widetilde{\overline{\eta }}^{N}_{(s)}\widetilde{\gamma }^{i}\widetilde{\eta }^{N}_{(s)} & \: \: \:  & K^{i}_{(t)MN}=\widetilde{\overline{\eta }}^{N}_{(t)}\widetilde{\gamma }^{i}\widetilde{\eta }^{N}_{(t)}
\end{eqnarray*}
which enjoy the useful properties
\begin{eqnarray*}
J_{PM}\: J_{PN}=J_{MP}\: J_{NP}=\delta _{MN} &  & \\
K^{i}_{(s)MP}\: J_{PN}=K^{i}_{(t)MN} &  & 
\end{eqnarray*}

Before we can extract the \( \kappa  \) symmetry transformations for the \( AdS_{3}\otimes S_{3}\otimes T^{4} \)
fields, it is important to realize that \( \widetilde{\eta }^{N}_{(s)} \)and
\( \widetilde{\eta }^{N}_{(t)} \) are \emph{not} linearly independent since
they both satisfy eq. (\ref{susy_const_1}) and there are at most 4 spinors
which are linearly independent over \( C \). Both of them satisfy the Fierz
identity
\begin{equation}
\label{fierz_eta-eta}
\widetilde{\eta }^{N}_{(s)}\widetilde{\overline{\eta }}^{N}_{(s)}=\widetilde{\eta }^{N}_{(t)}\widetilde{\overline{\eta }}^{N}_{(t)}=1_{2}\otimes \frac{1+\rho \overline{\gamma }_{5}}{2}
\end{equation}
since both \( \widetilde{\eta }^{N}_{(s)} \) and \( \widetilde{\eta }^{N}_{(t)} \)
are orthonormal as follows from eq. (\ref{norm-cond-eta}).

We are now ready to insert eq.s (\ref{ansatz-epsilon},\ref{ansatz-kappa})
into eq. (\ref{kappa-symmetry}) in order to get the relation between the parameters
\( \epsilon _{N} \) of the supersymmetry variation and the parameters \( \kappa _{N} \)
:
\begin{eqnarray}
\epsilon ^{(+)}_{N}=\Pi ^{a}_{\alpha }\: \gamma _{a}\kappa _{N}^{(+)\alpha }-i\: \Pi ^{i}_{\alpha }\: K^{i}_{(s)NM}\: \kappa ^{(+)\alpha }_{M} &  & \label{kappa-symmetry-3D} \\
\epsilon ^{(-)}_{N}=\Pi ^{a}_{\alpha }\: \gamma _{a}\kappa _{N}^{(-)\alpha }-i\: \Pi ^{i}_{\alpha }\: K^{i}_{(t)NM}\: \kappa ^{(-)\alpha }_{M} &  & \label{kappa-symmetry-3D-b} 
\end{eqnarray}
and into eq. (\ref{vincolo-su-kappa}) in order to get the constraints on the
\( \kappa  \) parameters:
\begin{equation}
\label{vincolo-kappa-symmetry-3D}
\kappa ^{(+)}_{N\alpha }+e^{i\left( \phi _{(+)}-\phi _{(-)}\right) }J_{NM}\kappa ^{(-)}_{M\alpha }=e^{i\left( \phi +2\phi _{(+)}\right) }\: \epsilon _{\alpha \beta }\left( \kappa ^{(+)}_{N\beta }+e^{i\left( \phi _{(-)}-\phi _{(+)}\right) }J_{NM}\kappa ^{(-)}_{M\beta }\right) 
\end{equation}

We could have proceeded directly in \( D=3 \) and imposed that eq. (\ref{S-primo-ordine})
were invariant under the susy transformation rules (\ref{susy-var-3D-1}-\ref{susy-var-3D-4})
with properly restricted parameters \( \epsilon ^{(\pm )}_{N} \) , then starting
from an ansatz like eq.s (\ref{kappa-symmetry-3D}-\ref{kappa-symmetry-3D-b})
with the coefficients undetermined and imposing the cancellation of the terms
proportional to \( \Pi ^{a}\Pi ^{i} \) we would have fixed these undetermined
coefficients and found a constraint equivalent to (\ref{vincolo-kappa-symmetry-3D}).

\section{Fixing the \protect\( \kappa \protect \) symmetry with the supersolvable projector.\label{app_kappa-symm}}

In this appendix we want to show that it is possible to fix the \( \kappa  \)
symmetry using the supersolvable algebra and find which states can be described
in this gauge: this is completely analogous to what happens in the more usual
GS formulation of the superstring in \( D=10 \) where only the states with
\( p^{+}\neq 0 \) can be described.

As a first step in this direction we introduce the projectors
\begin{eqnarray*}
P=\frac{1}{\sqrt{2}}\gamma ^{0}\gamma ^{-}=\frac{1+i\gamma ^{2}}{2},Q=\frac{1}{\sqrt{2}}\gamma ^{0}\gamma ^{+}=\frac{1+i\gamma ^{2}}{2} &  & 
\end{eqnarray*}
and we notice that the request of fixing the \( \kappa  \) symmetry by using
the supersolvable algebra is equivalent to showing the possibility of setting
\( P\psi ^{(+)}=Q\psi ^{(-)}=0 \) , and this can be reformulated as the fact
we can find a solution for \( \kappa ^{(\pm )}_{N\alpha } \) for any \( P\epsilon ^{(+)} \)
and \( Q\epsilon ^{(-)} \) .

In order to show this we need introduce some notation in order to be able to
write the following expression in a more concise way; we define therefore
\begin{eqnarray*}
C=\cos (\phi +2\phi _{(+)})=-\frac{\Re (ee^{i\phi })}{\left| e\right| } & , & S=\sin (\phi +2\phi _{(+)})=-\frac{\Im (ee^{i\phi })}{\left| e\right| }\\
\lambda ^{(-)}_{N\alpha }=\kappa ^{(+)}_{M\alpha }J_{MN} & , & \eta ^{(-)}_{N\alpha }=\epsilon ^{(+)}_{M\alpha }J_{MN}
\end{eqnarray*}
so that eq. (\ref{vincolo-kappa-symmetry-3D}) implies that not all the components
of \( \kappa ^{(-)}_{\alpha } \) and \( \lambda ^{(-)}_{\alpha } \)are independent,
i.e. (from now on we drop all the \( ^{(-)} \) in order to avoid to burden(??)
the notation too much)
\begin{eqnarray}
-\kappa _{N1}=C\kappa _{N0}+S\lambda _{N0} &  & \nonumber \\
-\lambda _{N1}=-C\lambda _{N0}+S\kappa _{N0} &  & \label{kappa-symm-const-explicit} 
\end{eqnarray}

The equations which are to be solved and relate \( P\eta  \) and \( Q\epsilon  \)
to \( \kappa  \) and \( \lambda  \) can be derived from (\ref{kappa-symmetry-3D},\ref{kappa-symmetry-3D-b})
read
\begin{eqnarray}
Q\epsilon _{N}=\Pi ^{p}_{\alpha }\: \gamma _{p}P\kappa _{N}^{\alpha }+\left( \Pi ^{2}_{\alpha }\: \gamma _{2}-i\: \Pi ^{i}_{\alpha }\: K^{i}_{(t)MN}\right) \: Q\kappa ^{\alpha }_{M} &  & \label{parametri-a-kappa-in-SS-0} \\
P\eta _{N}=\Pi ^{p}_{\alpha }\: \gamma _{p}Q\lambda _{N}^{\alpha }+\left( \Pi ^{2}_{\alpha }\: \gamma _{2}-i\: \Pi ^{i}_{\alpha }\: K^{i}_{(t)MN}\right) \: P\lambda ^{\alpha }_{M} &  & \label{parametri-a-kappa-in-SS} 
\end{eqnarray}
and when we set \( P\kappa =Q\lambda =0 \) and utilize the constraints (\ref{kappa-symm-const-explicit})
they can be rewritten as
\[
\left( \begin{array}{c}
Q\epsilon _{N}\\
P\lambda _{N}
\end{array}\right) =\left( \begin{array}{cc}
\left( \Pi ^{p}_{0}\: \gamma _{p}+C\: \Pi ^{p}_{1}\: \gamma _{p}\right) Q\: \delta _{NM} & S\: \left( \Pi ^{2}_{1}\: \gamma _{2}\: \delta _{NM}-i\: \Pi ^{i}_{\alpha }\: K^{i}_{(t)MN}\right) P\\
S\: \left( \Pi ^{2}_{1}\: \gamma _{2}\: \delta _{NM}-i\: \Pi ^{i}_{\alpha }\: K^{i}_{(t)MN}\right) Q & \left( \Pi ^{p}_{0}\: \gamma _{p}-C\: \Pi ^{p}_{1}\: \gamma _{p}\right) P\: \delta _{NM}
\end{array}\right) \left( \begin{array}{c}
P\kappa _{M0}\\
Q\lambda _{M0}
\end{array}\right) \]
When we compute the determinant of the matrix entering the previous expression
on the subspace where \( Q=P=1 \)and we require it to be different from zero
we get the constraint on the states which can be described by this gauge
\begin{equation}
\label{stati-che-si-descrivono-appF}
2\left( \Pi ^{+}_{0}-C\Pi ^{+}_{1}\right) \left( \Pi ^{-}_{0}+C\Pi ^{-}_{1}\right) +S^{2}\left( \Pi _{1}^{2}\Pi _{21}+\Pi _{1}^{i}\Pi _{i1}\right) \neq 0
\end{equation}

We notice that the request \( P\kappa =Q\lambda =0 \) is not restrictive at
all and whichever solution to the eq.s (\ref{parametri-a-kappa-in-SS-0}-\ref{parametri-a-kappa-in-SS})
implies eq. (\ref{stati-che-si-descrivono-appF}).

\section{Another useful expression for the pure RR action.\label{action_D6_fermion}}

We want now find a different expression for the action (\ref{The-Action}) where
we make use of 6D pseudo-Majorana Weyl spinors (\cite{ParkRey} ). To this purpose
we construct the 6D gamma matrices as
\begin{eqnarray*}
\Gamma _{a}=\gamma _{a}\otimes 1\otimes \sigma _{1\: },\:  & \Gamma _{i}=1\otimes \gamma _{i}\otimes \sigma _{2\: }, & \: {\cal C}=C\otimes C\otimes \sigma _{1}\\
 & \Gamma _{7}=-\Gamma _{0}\ldots \Gamma _{5}=1\otimes 1\otimes \sigma _{3} & 
\end{eqnarray*}
which satisfy the properties

\[
{\cal C}^{T}={\cal C}\: \: ,\: \: \Gamma ^{T}_{\bullet }=-{\cal C}\Gamma _{\bullet }{\cal C}\]

To proceed further we need the explicit expression of the \( S_{3} \) spinors
which satisfy eq. (\ref{kill_spin_s3_pure}):
\[
\eta ^{p}_{(s)}=\frac{1}{\sqrt{1-y^{2}}}\left( 1+s\rho \: y^{i}\: \gamma _{i}\right) \epsilon ^{p}\]
where \( \epsilon ^{p} \) is a set of 2 constant spinors normalized as \( \overline{\epsilon }_{q}\epsilon ^{p}=\delta ^{p}_{q} \).
This can be derived as in (\cite{Mio_all} ).

With these definitions we can verify, using the second equation of (\ref{kill_spin_decomp}),
that the following 6D spinors

\[
\Theta _{p}=(\theta ^{(+)}_{N}+\theta ^{(-)}_{N})\otimes \epsilon ^{r}\: H^{N}_{rp}\otimes \xi \]
where \( \xi =\left( \begin{array}{c}
i\\
0
\end{array}\right)  \) are a couple of pseudo-Majorana Weyl spinors; explicitly we have
\[
\Theta _{cp}\equiv {\cal C}\overline{\Theta }^{T}_{p}=\epsilon _{pq}\: \Theta _{q}\: \: \: ,\: \: \: \Gamma _{7}\Theta _{p}=+\Theta _{p}\]
We notice that in performing the sum \( \theta ^{(+)}_{N}+\theta ^{(-)}_{N} \)
we do not loose any d.o.f since \( \sigma _{3}\theta ^{(\pm )}_{N}=\pm \theta ^{(\pm )}_{N} \).

With the further change of bosonic variables (\( y^{2}=-\delta _{ij}y^{i}y^{j} \)
)
\begin{eqnarray*}
Y^{2} & = & r\frac{1+y^{2}}{1-y^{2}}\\
Y^{i} & = & (-s\rho )\: r\frac{2y^{i}}{1-y^{2}}
\end{eqnarray*}
we can now easily rewrite the action as follows (\( (Y)^{2}=-\left( Y^{2}\right) ^{2}-\left( Y^{i}\right) ^{2} \)
) 
\begin{eqnarray}
 & S=\int _{\Sigma _{2}}d^{2}\xi \, \sqrt{-g}\, g^{\alpha \beta }\frac{1}{2}\times  & \nonumber \\
 & \left\{ -2|e|^{2}\: \left( Y\right) ^{2}\: \left[ \partial _{\alpha }x^{+}+\frac{i}{2}\overline{\Theta }^{p}\Gamma ^{+}\partial _{\alpha }\Theta _{p}\right] \, \left[ \partial _{\beta }x^{-}+\frac{i}{2}\overline{\Theta }^{p}\Gamma ^{-}\partial _{\beta }\Theta _{p}\right] -\frac{1}{|e|^{2}}\delta _{ut}\frac{\partial _{\alpha }Y^{t}\: \partial _{\beta }Y^{u}}{\left( Y\right) ^{2}}-\delta _{rs\: }\partial _{\alpha }z^{r}\: \partial _{\beta }z^{s}\right\}  & \nonumber \\
 & -|e|\Im e\: \left( Y\right) ^{2}\: \left[ dx^{+}+\frac{i}{2}\overline{\Theta }^{p}\Gamma ^{+}d\Theta _{p}\right] \wedge \left[ dx^{-}+\frac{i}{2}\overline{\Theta }^{p}\Gamma ^{-}d\Theta _{p}\right] +\frac{e}{2\mid e\mid }Y^{t}id\overline{\Theta }^{p}\Gamma _{t}d\Theta _{p}+S_{WZWN}\label{action_6D} 
\end{eqnarray}
where \( t,u=2,\ldots ,5 \) can be interpreted as the directions transverse
to the \( (p,q) \) D1 brane generating this background in the near horizon
limit when we forget the \( T^{4} \) \( \sigma  \) model.


\begin{thebibliography}{}
\bibitem{Maldacena}J. Maldacena, The Large \( N \) Limit of Superconformal Field Theory and Supergravity,
hep-th/9711200
\bibitem{GubserKlenanovPolyako}S.S. Gubser, I.R. Klebanov and A.M. Polyakov, Gauge Theory Correlators from
Non-Critical String Theory, hep-th/9802109
\bibitem{Witten}E. Witten, Anti de Sitter Space and Holography, hep-th/9802150
\bibitem{Mio_ads}I. Pesando, A \( \kappa  \) Gauge Fixed Type IIB Superstring Action on \( AdS_{5}\times S_{5} \),
hep-th/9808020
\bibitem{Kall_ads}R. Kallosh and J. Rahmfeld, The GS String Action on \( AdS_{5}\times S_{5} \),
hep-th/9808038
\bibitem{Mio_all}I. Pesando, All Roads lead to Rome: Supersolvable and Supercosets, hep-th/9808146
\bibitem{MatsaevTseytlin}R.R Metsaev and A.A. Tseytlin, Type IIB Superstring Action in \( AdS_{5}\times S_{5} \)
Background, hepth/9805028; 
\bibitem{Kallosh}R. Kallosh, J. Rahmfeld and A. Rajaraman, Near Horizon Superspace, hep-th/9805217
\bibitem{Kall-fixing}R. Kallosh, Superconformal Actions in Killing Gauge, hep-th/9807206
\bibitem{kall-raj-vuoti}R. Kallosh and A. Rajaraman, Vacua of M-theory and String Theory, hep-th/9805041
\bibitem{kall_ts-simpl}R. Kallosh and A.K Tseytlin, Simplifying Superstring Action on \( AdS_{5}\times S_{5} \),
hep-th/9808088
\bibitem{4-punti}D.Z. Freedman, S.D. Mathur, A. Matusis and L. Rastelli, Comments on 4-point
functions in CFT/AdS correspondence, hep-th/9808006
\bibitem{4-punti-corretta}J.H Brodie and M. Gutperle, String corrections to four point functions in the
CFT/AdS correspondence, hep-th/9809067
\bibitem{ads3-sei}A. Giveon, D. Kutazov and N. Seiberg, Comments on String Theory on \( AdS_{3} \)
, hep-th/9806194
\bibitem{IIB}L. Castellani and I. Pesando, The Complete Superspace Action of Chiral \( D=10 \)
\( N=2 \) Supergravity, Int. Jou. Mod. Phys. A8 (1993) 1125;\\
L. Castellani, Chiral \( D=10 \),\( N=2 \) Supergravity on the Group Manifold:
1. Free Differential Algebra and Solution of Bianchi Identities ,Nucl. Phys.
B294 (1987) 877
\bibitem{GruppoToGH}L. Castellani, A. Ceresole, R. D'Auria, S. Ferrara, P. Fre' and M. Trigiante,
\( G/H \) M-branes and \( AdS_{p+2} \) Geometries, hep-th/9803039
\bibitem{gruppoToOsp}G. Dall'Agata, D. Fabbri, C. Fraser, P. Fre', P. Termonia and M. Trigiante,
The \( Osp(8|4) \) Singleton Action from the Supermembrane, hep-th/9807115
\bibitem{GHMNT}M.T. Grisaru, P. Howe, L. Mezincescu, B.E.W. Nilsson and P.K. Townsend, \( N=2 \)
Superstrings in a Supergravity Background, Phys. Lett. 162B (1985) 116.
\bibitem{PST}P. Pasti, D. Sorokin and M.Tonin, On gauge-fixed superbrane actions in \( AdS \)
superbackgrounds, hep-th/9809213
\bibitem{ParkRey}J. Rahmfeld and A. Rajaraman, The GS String Action on \( AdS_{3}\times S^{3} \)with
Ramond-Ramond Charge, hep-th/9809164\\
J. Park and S. Rey, Green-Schwarz Superstring on \( AdS_{3}\times S^{3} \),
hep-th/9812062
\end{thebibliography}
\end{document}